# Paired Conditional Generative Adversarial Network for Highly Accelerated Liver 4D MRI

Di Xu[1], Xin Miao[2], Hengjie Liu[3], Jessica E. Scholey[1], Wensha Yang[1], Mary Feng[1], Michael Ohliger[1], Hui Lin[1], Yi Lao[3], Yang Yang[4] and Ke Sheng[1, *]

1 Radiation Oncology, University of California, San Francisco, 505 Parnassus Ave, San Francisco, CA 94143
2 Siemens Healthineers, 1 Federal St, Boston, MA 02110
3 Radiation Oncology, University of California, Los Angeles, 200 Medical Plaza, Los Angeles, CA 90095
4 Radiology, University of California, San Francisco, 505 Parnassus Ave, San Francisco, CA 94143
Corresponding author: ke.sheng@ucsf.edu

## Abstract

**Purpose**: 4D MRI with high spatiotemporal resolution is desired for image-guided liver radiotherapy. Acquiring densely sampling k-space data is time-consuming. Accelerated acquisition with sparse samples is desirable but often causes degraded image quality or long reconstruction time. We propose the Reconstruct Paired Conditional Generative Adversarial Network (Re-Con-GAN) to shorten the 4D MRI reconstruction time while maintaining the reconstruction quality.

**Methods**: Patients who underwent free-breathing liver 4D MRI were included in the study. Fully- and retrospectively under-sampled data at 3, 6 and 10 times (3x, 6x and 10x) were first reconstructed using the nuFFT algorithm. Re-Con-GAN then trained input and output in pairs. Three types of networks, ResNet9, UNet and reconstruction swin transformer, were explored as generators. PatchGAN was selected as the discriminator. Re-Con-GAN processed the data (3D+t) as temporal slices (2D+t). A total of 48 patients with 12332 temporal slices were split into training (37 patients with 10721 slices) and test (11 patients with 1611 slices). Compressed sensing (CS) reconstruction with spatiotemporal sparsity constraint was used as a benchmark. Reconstructed image quality was further evaluated with a liver gross tumor volume (GTV) localization task using Mask-RCNN trained from a separate 3D static liver MRI dataset (70 patients; 103 GTV contours).

**Results**: Re-Con-GAN consistently achieved comparable/better PSNR, SSIM, and RMSE scores compared to CS/UNet models. The inference time of Re-Con-GAN, UNet and CS are 0.15s, 0.16s, and 120s. The GTV detection task showed that Re-Con-GAN and CS, compared to UNet, better improved the dice score (3x Re-Con-GAN 80.98%; 3x CS 80.74%; 3x UNet 79.88%) of unprocessed under-sampled images (3x 69.61%).

**Conclusion**: A generative network with adversarial training is proposed with promising and efficient reconstruction results demonstrated on an in-house dataset. The rapid and qualitative reconstruction of 4D liver MR has the potential to facilitate online adaptive MR-guided radiotherapy for liver cancer.

## 1. Introduction

MRI has been increasingly adopted for image-guided liver radiation therapy (RT) owing to its superior soft tissue contrast compared to CT[1–3]. 4D MRI, which is a respiratory-resolved volumetric imaging technique, is especially powerful in quantifying tumor motion[4–6]. In a clinical planning workflow for free-breathing liver RT treatments, the liver tumor delineated in individual 4D MR images forms an internal target volume (ITV). Under or over-estimating ITV can cause tumor underdose or normal tissue injury during radiation.

4D MRI data is usually acquired in a continuous free-breathing scan followed by data sorting based on respiratory motion surrogates. One common option is a 3D golden angle stack-of-stars sequence with self-navigation to acquire 4D MRI datasets[5]. In stack-of-stars acquisition, radial sampling is employed in the $kx - ky$ plane, which enables reduced motion sensitivity[7] and incoherent k-space under-sampling if acceleration is desired[8]. Cartesian sampling is used in the kz dimension, which allows for a flexible selection of volumetric coverage/slice resolution[9]. However, this technique has a few limitations. Scan time is usually long (8-10 min[5]), and slice resolution is often sacrificed to maintain sufficient volumetric coverage and in-plane resolution, increasing the inaccuracy of small malignancy contouring[9]. Streak artifacts caused by under-sampling, sampling trajectory deviation, or nonuniform k space coverage can be challenging to mitigate with conventional constrained reconstruction. Parallel imaging[10,11] and compressed sensing[12–14] have been employed to accelerate both static and dynamic MRI.

Recent deep learning (DL) advances have offered a data-driven approach for 4D MRI reconstruction. In contrast to the model-based CS reconstruction, DL learns reconstruction mapping from the rich information in the training data representation, matching or exceeding the CS quality and is significantly faster[15–17]. Previous works have explored 4D MRI reconstruction using customized convolutional neural networks (CNNs), recurrent neural networks (RNNs), and Transformers. For instance, Schlemper et al. proposed integrating K nearest neighbor (KNN) enabled temporal data-sharing mechanism into a cascade 3D CNN architecture for 4D MRI reconstruction trained with L2 loss objective[16]. Their network demonstrates the capability of 2D frame recovery but is suboptimal in capturing complex dynamic relationships. Adapting from the UNet architecture, Dracula[18] and Moivenet[19] proposed methods to accelerate 4D MR reconstruction. However, there is still room for improvement in the reduction of reconstruction time (Dracula at 28 s and Moivenet at 0.69 s). Moreover, Huang et al. introduced a motion-guided framework using RNN-inspired Conv-GRU for initial 2D frame reconstruction and U-FlowNet for motion estimation in the optical flow field. The overall architecture is trained with regularized L1 loss[20]. Their pipeline reconstructed a cardiac dataset 5 and 8 times accelerated (5x and 8x), but detail loss was evident at a high acceleration ratio. Additionally, their RNN-based architecture confines its tensor processing to be sequentially frame-by-frame and takes ~5 s for inference of a volume, which limits its application for MRI-guided real-time interventions[21]. Lately, Xu et al. designed a 2D CNN-assisted Reconstruction Swin Transformers (RST), a variant of Video Swin Transformers[22], supervised with a combination cost function of L1, peak signal-to-noise ratio (PSNR), and multi-scale structure similarity index measurement (MS-SSIM)[23] for 4D MRI reconstruction and validates the algorithm on a 9x accelerated cardiac dataset[21]. The RST architecture showed promising results in dynamic relationship learning and faster inference (<1s), but struggled to retain finer spatial anatomies[24–26].

It is well known that the success of neural network (NN) training, apart from exploring various architectures, hinges on the loss function design[27,28]. However, designing an effective loss function that encourages NNs to precisely converge towards the target often requires balancing conflicting constraints such as sharpness vs. streak artifacts reduction. Generative Adversarial Networks (GANs)[29] took an alternative approach – rather than explicitly specifying all components of the loss function, the discriminator network implicitly guides the generator loss reduction by distinguishing real from synthesized images. This adversarial dynamic can lead to outputs that are closer to the ground truth target. Inconsistent and blurry predictions are discriminated against in GANs. Additionally, since GANs only require the generator model at inference time, the adversarial training process does not add computational burden during reconstruction. This makes real-time 4D MRI reconstruction more tractable[29].

Several previous studies have demonstrated superior MRI reconstruction using GANs. For Cartesian sampling, Yang et al.[30] demonstrated that their UNet-based conditional GAN could provide better reconstruction with preserved perceptual imaging details than non-adversarial CNN methods on 3D T1-weighted brain and cardiac MRI dataset. Mardani et al.[31] built a least squares conditional GAN, demonstrating competitive performance in pediatric contrast-enhanced 3D MRI reconstruction. For non-Cartesian sampling, Liu et al.[32] presented a robust performance by cycle-GAN trained with varying under-sampling patterns on 3D golden-angle radial sampled liver imaging. Gao et al.[33] also demonstrated the feasibility of using a conditional GAN framework for 3D stack-of-radial Liver MRI reconstruction. However, the previous work has not explored the capability of GANs in 4D MRI temporal profiling and reconstruction.

The current work explores the feasibility of using GANs for 4D MRI reconstruction. We have developed a novel architecture termed Reconstruct paired Conditional GAN (Re-Con-GAN), specifically for 4D MRI reconstruction. The proposed framework is designed to learn 2D+time image series from under-sampled data. Experiments on an in-house 4D liver MRI dataset demonstrate the superior performance of Re-Con-GAN compared to conventional compressed sensing and supervised deep learning reconstruction models. To further validate the robustness of Re-Con-GAN's reconstructed images, we evaluate the impact on downstream tasks of liver tumor detection and segmentation using a Mask R-CNN[34] pipeline. The rest of the manuscript is organized as follows: Section 2 elaborates on data cohort, Re-Con-GAN framework, baseline algorithms, and model evaluation; Section 3 summarizes the experimental results; and Section 4, along with Section 5, discusses and concludes the current work.

## 2. Materials and Methods

### 2.1 4D MR Data Cohort

The study was approved by the local Institutional Review Board at UCSF (#14-15452). 48 patients were scanned on a 3T MRI scanner (MAGNETOM Vida, Siemens Healthcare, Erlangen, Germany) after injection of hepatobiliary contrast (gadoxetic acid; Eovist, Bayer) for each patient. A prototype free-breathing T1-weighted volumetric golden angle stack-of-stars sequence was used for 4D MRI acquisition. The scanning parameters were - TE=1.5 ms, TR=3 ms, matrix size = 288x288, FOV = 374 mm x 374 mm, in-plane resolution=1.3 mm × 1.3 mm, slice thickness=3 mm, radial views (RV) per partition=3000, number of slices or partitions = 64-75, acquisition time = 8-10 min. The pulse sequence ran continuously over multiple respiratory cycles. Images reconstructed from the entire space data of 3000 radial spokes (RV-3000) were treated as the fully sampled ground truth reference (Based on Nyquist sampling theorem, fully sampled radial images require sampling points $\times \frac{\pi}{2}$ spokes, resulting in 452 spokes for a matrix size of $288 \times 288$. After motion binning, each of the 8 bins has, on average, 375 spokes with RV3000, which is close to 452 spokes and could well preserve imaging quality). Retrospective under-sampling was performed by keeping the first 1000, 500, and 300 spokes from the 3000 spokes, respectively, corresponding to acceleration rates of 3x, 6x, and 10x. For initial image reconstruction, data sorting based on a self-gating signal was performed to divide the continuously acquired k-space data into 8 respiratory phases. nonuniform fast Fourier transform (nuFFT) algorithm was applied to reconstruct each phase individually.

Only regular breathers (48 patients) were included in the current project. Breathing regularity was quantified using the self-gating signal waveform[32,35,36]. The peak-to-trough range and mid-level amplitude (A), i.e., (peak-A + trough-A)/2, were calculated for each respiratory cycle. The average mid-level amplitude across all respiratory cycles normalized with the average peak-to-trough range was used as the regularity measurement. Patients with a score greater than 20% were classified as irregular breathers and excluded.

To augment the sample size, we organized the 48 3D+t data as 12332 2D+t images with images from an individual patient sorted in one subset. The data was split into training (37 patients with 10721 2D+t images) and testing (11 patients with 1611 2D+t images), where patients with various profiles (body mass index and breathing regularity score) are balanced in each split. The images were resized to $256 \times 256$ and normalized using Z-score normalization. Data augmentation was employed, including random rotation, flipping, and cropping.

## 2.2 Dynamic MR Paired Conditional GANs

### *2.2.1 Network Architecture*

The architecture of GANs can be designed in unconditional or conditional settings. Unconditional GANs learn a mapping from a random noise vector $z$ to output image $y$, $G: z \rightarrow y$, whereas conditional GANs (cGANs) learn a mapping from an observed image $x$ as well as a random noise vector $z$ to output image $y$, $G: \{x, z\} \rightarrow y$. cGANs can be further classified into paired and unpaired architectures. Paired cGANs learn a one-to-one mapping of input to output, while unpaired cGANs only conduct domain-level supervision with input and output randomly selected from its domain data corpus. The current work employed a cGAN structure to perform the image-domain 4D MRI reconstruction as a paired image-to-image translation task.

Paired cGANs consist of two major components – generator $G$ and discriminator $D$. On the one hand, $G$ is trained to generate a “fake” reconstructed image series that cannot be differentiated from their corresponding "real" fully sampled ground truth (GT) image series by $D$. On the other hand, $D$ is trained to classify between "fake", $G$ synthesized image series $\hat{G}(x)$ and $x$, and "real", fully sampled image series $y$ and $x$, tuples. In cGANs, both $G$ and $D$ can observe input under-sampled image series. Details of the discriminator training workflow is diagrammed in **Figure 1**.

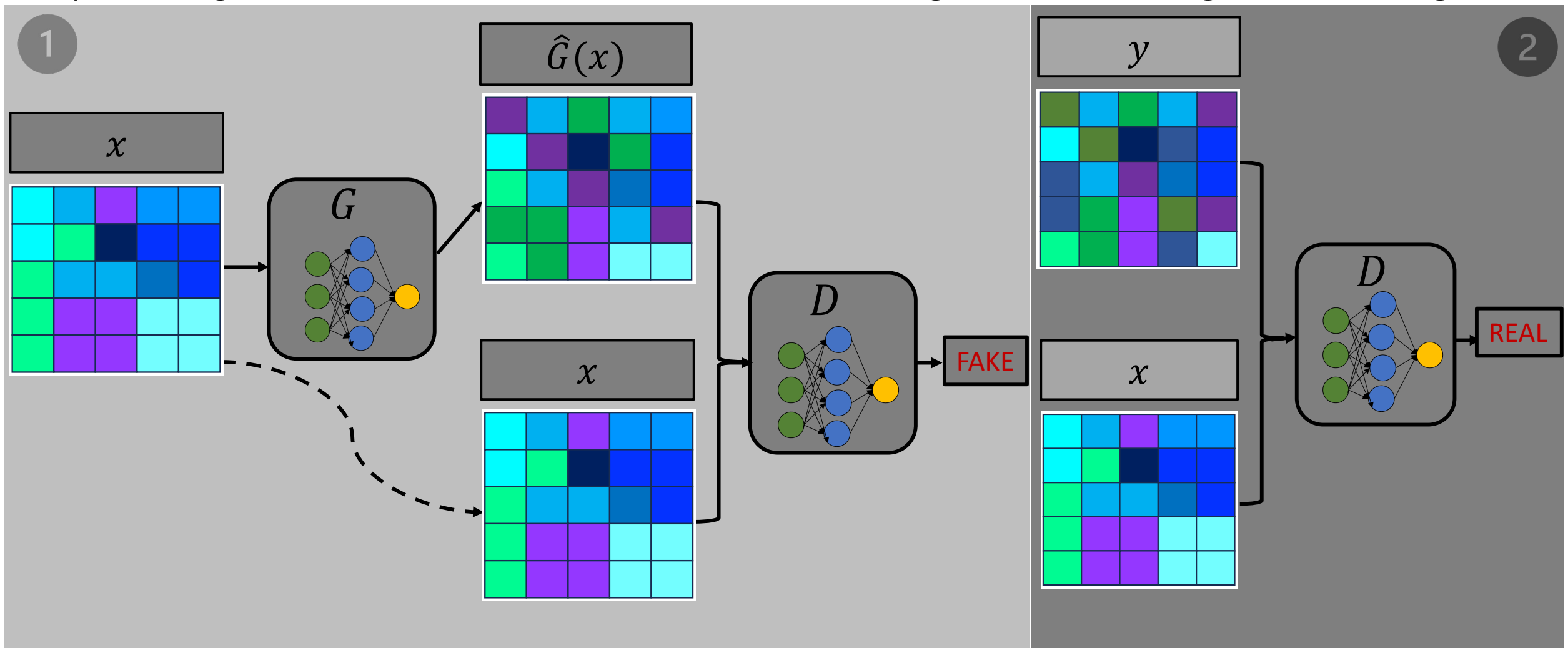

**Figure 1**: Training cGANs Discriminator $D$ to map input images $x$ to target domain $y$. $D$ is designed to learn to distinguish between fake ($\hat{G}(x)$; synthesized by Generator $G$) and real ($y$; GT).

Our design of the generator and discriminator is improved from Isola et al.[29]. Re-Con-GAN is a versatile architecture with plug-and-playable generator, discriminator, and loss objective. A couple of examples for each sub-component are experimented with and demonstrated in this paper. Details of the model architectures can be visualized in **Figure 2** and are elaborated as follows.

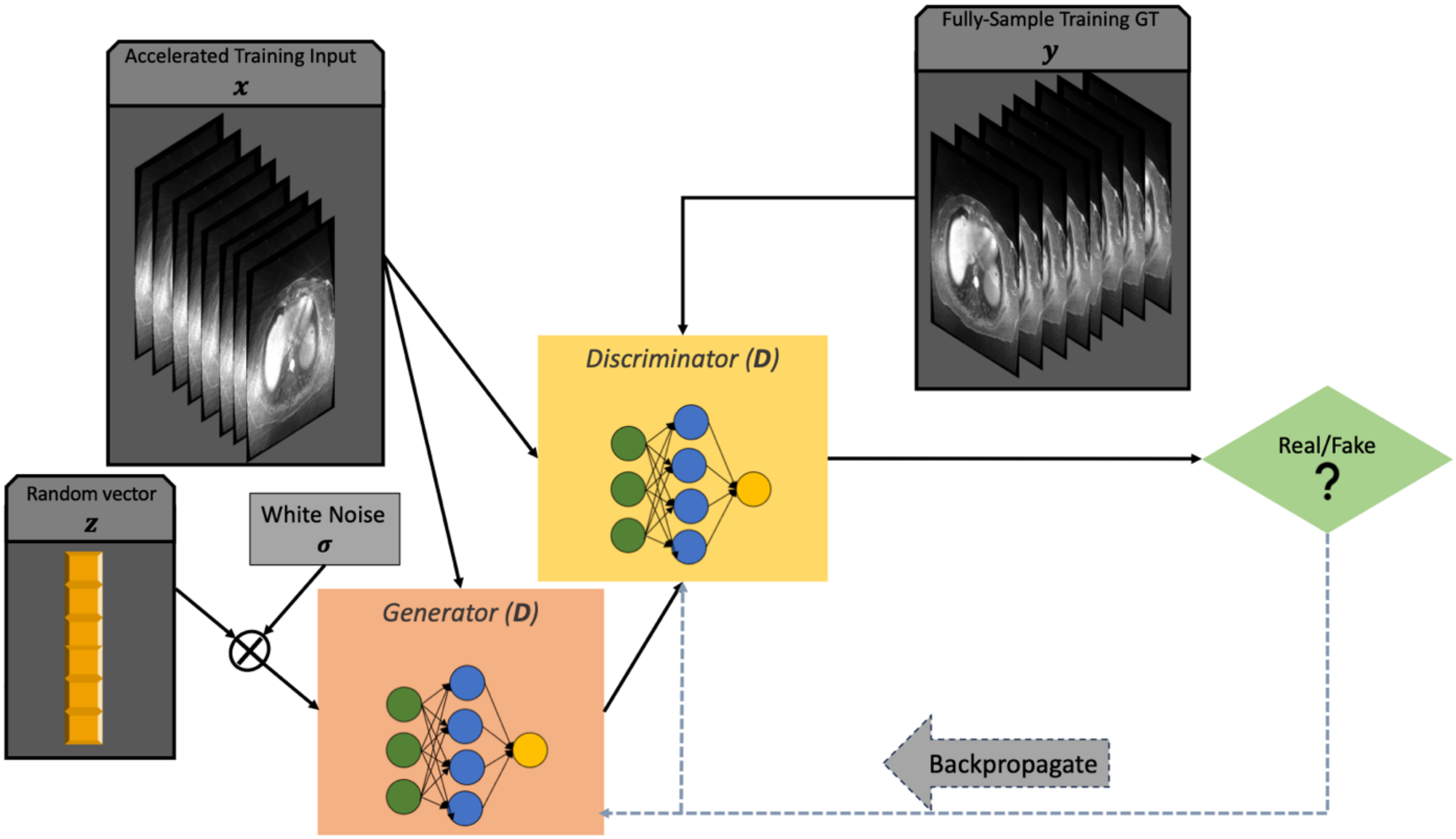

**Figure 2**: The design of our proposed Re-Con-GAN. The input to Re-Con-GAN is a random noise vector $z$ and under-sampled image series $x$. The supervision is a fully sampled image series $y$. The network is trained by combining generator $G$ and discriminator $D$.

#### 2.2.1.1 Generator

The defined task for the generator of 4D MRI reconstruction is to map a low-resolution input grid with noise and artifacts to a high-resolution noise- and artifact-reduced output grid. The input and output image series differ in morphological details and a few surface structures. But they intrinsically both render the same underlying general structures. Several previous DL approaches solve the 4D MRI reconstruction problem with encoder-to-decoder NNs[16,21]. Such architectures conduct a series of down-sampling operations until they reach the bottleneck layer, and then reverse the operations to up-sampling to gradually recover the dimension of the feature map from that of input. The edge of the encoder-to-decoder architecture is that numerous low-level information shared between input and output is revealed per progressively down-sampling while the difference between output and input is parameterized with the up-sampling layers.

**Figure 3** shows three encoder-to-decoder network designs, including ResNet9[37], UNet[38] and RST[21]. Considering that the data is organized as 2D+t image series, the three NNs were all convolved in the 3D domain with the t dimension of the data propagating through channel dimension in networks (3D ResNet9, 3D UNet, and 3D RST).

**3D ResNet9**: The 3D ResNet9 is adopted from Johnson et al.[37]. Let $c7s1-k$ denote a residual block consisting of a $7 \times 7$ convolutional layer with $k$ number of filters and stride of 1, Instance normalization and ReLU operations. $dk$ denotes a residual block consisting of a $3 \times 3$ convolutional layer with $k$ number of filters and stride of 2, instance normalization and ReLU operations. $Rk$ denotes a residual block consisting of two of $3 \times 3$ convolutional layers with the same $k$ number of filters and stride of 2, instance normalization, and ReLU operations. $uk$ denotes a fractional-stride residual block consisting of two of $3 \times 3$ convolutional layers with the same $k$ number of filters and stride of $\frac{1}{2}$, instance normalization, and ReLU operations. Reflection padding is used per convolution to reduce artifacts. The 3D ResNet9 is structured as Equation (1).

$$c7s1-64, d128, d256, R256, R256, R256, R256, R256, R256, R256, R256, R256, R256, R256, u64, c7s1-8 \quad (1)$$

Where $c7s1-8$ is used to map the prediction to the expected number of output channels (8 in the current paper).

**3D UNet**: The 3D UNet structure follows the design of Isola et al.[29]. Let $\mathcal{C}k$ denote a UNet block consisting sequentially of a $4 \times 4$ convolutional layer with $k$ number of filters and stride of 2, batch normalization and ReLU operations. Let $\mathcal{CD}k$ denote a UNet block consisting sequentially of a $4 \times 4$ convolutional layer with $k$ number of filter and stride of 2, batch normalization, dropout at 50%, and ReLU operations. Convolutions in the encoder stage down-sample by a factor of 2 at each block, whereas those in the decoder stage up-sample by a factor of 2. The 3D UNet is structured as Equation (2).

$$\begin{gathered} encoder: \mathcal{C}64, \mathcal{C}128, \mathcal{C}256, \mathcal{C}512, \mathcal{C}512, \mathcal{C}512, \mathcal{C}512, \mathcal{C}512 \\ decoder: \mathcal{CD}512, \mathcal{CD}1024, \mathcal{CD}1024, \mathcal{CD}1024, \mathcal{CD}1024, \mathcal{CD}512, \mathcal{CD}256, \mathcal{CD}64, \mathcal{CD}8 \end{gathered} \quad (2)$$

Where $\mathcal{CD}8$ is used to map the prediction to the expected number of output channels (8 in the current paper).

**3D RST**: The 3D RST structure follows the design of Xu et al.[21]. RST-Tiny (RST-T) is employed in the current work. Let $\mathcal{R}k$ denote an RST block with k number of filters consisting of a window multi-head self-attention layer (W-MSA) followed by a shifted window MSA (SW-MSA). A W-MSA unit consists sequentially of layer normalization, window self-attention, layer normalization, and multi-layer perception operations. The SW-MSA unit duplicates W-MSA, except that window self-attention is substituted with shifted window self-attention. $X \times \mathcal{R}k$ represents X number of identical $\mathcal{R}k$ blocks. The Encoder down-samples the feature by a factor of 2 at each block, whereas the decoder up-samples by a factor of 2. Skip connections between the encoder and decoder are not included to avoid GPU memory overflow at the decoding stage. The RST-T is structured as Equation (3).

$$\begin{gathered} encoder: 2 \times \mathcal{R}96, 2 \times \mathcal{R}192, 6 \times \mathcal{R}384, 2 \times \mathcal{R}768 \\ decoder: 2 \times \mathcal{R}768, 6 \times \mathcal{R}384, 2 \times \mathcal{R}192, 2 \times \mathcal{R}96, \mathcal{R}8 \end{gathered} \quad (3)$$

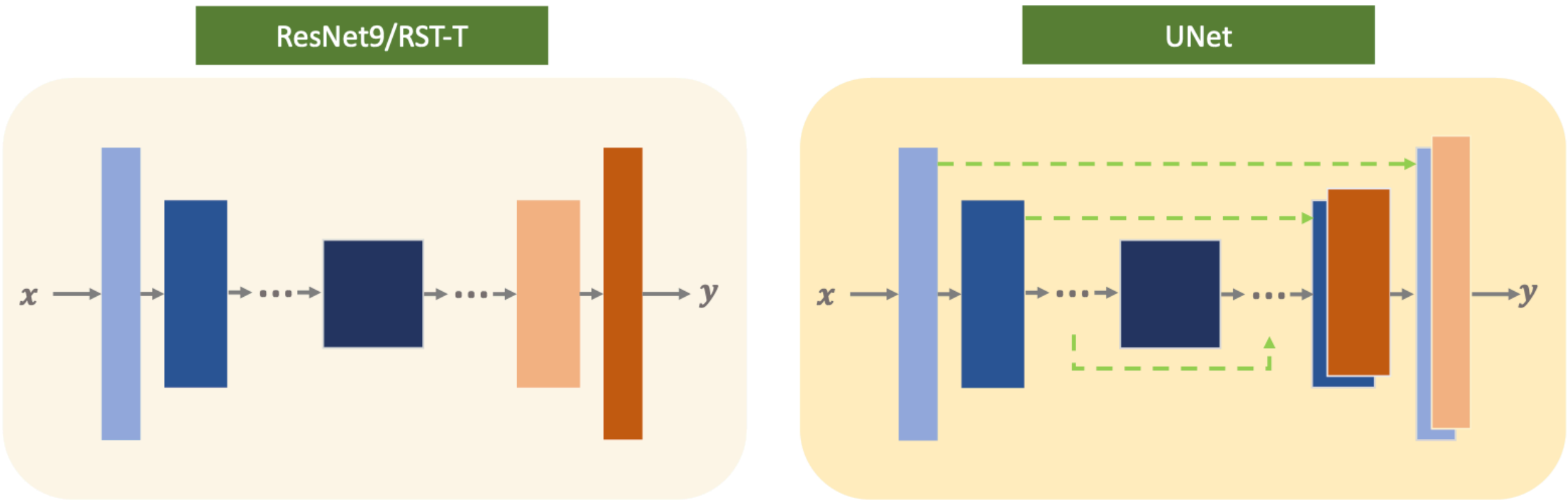

**Figure 3**: Three choices for the generator architecture.

#### 2.2.1.2 Discriminator

Following Isola et al.[29], a $N \times N$ PatchGAN is used as the discriminator in Re-Con-GAN. PatchGAN penalizes image series structure at the scale of patches. Specifically, PatchGAN works on identifying if each $N \times N \times t$ image series patch is a "real" or "fake" image series. The discriminator is run temporal-patch-based across the entire spatial dimension, averaging all the corresponding patch predictions from an image series to generate the final discrimination output. Assuming independence among pixels divided by more than a patch coverage, the PatchGAN discriminator essentially models the input image series as a Markov random field to understand the style/texture difference between "real" and "fake" image series[39,40]. The patch size $N$ is a tunable hyperparameter. As Isola et al.[37] discussed, smaller patch sizes have fewer parameters, thus running faster but potentially increasing tiling artifacts. Relatively larger patch sizes sacrifice the running speed but reduce the artifacts. $N$ is set as 70 across the entire current experiments. The $70 \times 70$ PatchGAN is structured as **Equation (4)** with $\mathcal{C}'k$ denoting a convolutional block consists sequentially of a $4 \times 4$ convolutional layer with $k$ number of filter and stride of 2, batch normalization and leaky ReLU operations.

$$\mathcal{C}'64, \mathcal{C}'128, \mathcal{C}'256, \mathcal{C}'512, \mathcal{C}'1 \tag{4}$$

After the last layer, a single-channeled smaller feature map is generated. An exception in Equation (4) is that Batch normalization is skipped in $\mathcal{C}'64$.

### *2.2.2 Loss Objective*

Earlier on, GANs formulated the discriminator as a classifier with a sigmoid cross entropy loss function, as shown in **Equation (5)**[29].

$$L_{cGAN(G,D)} = \mathbb{E}_{x,y}[\log D(x,y)] + \mathbb{E}_{x,z}[\log(1 - D(x, G(x,z))] \tag{5}$$

However, later studies show that the sigmoid cross entropy loss function is susceptible to vanishing gradient and has poor training stability during the learning process[41]. Therefore, least square GANs (LSGANs)[41] are proposed to remedy the issues. LSGANs modify the objective function as **Equation (6-7)**, penalizing sample feature maps based on their pixel distance to the corresponding decision boundary. In this way, more gradients are generated to update the generator.

$$L_{LScGAN(D)} = \frac{1}{2}\sum \mathbb{E}_{x,y}[(D(x,y) - b)^2] + \frac{1}{2}\sum \mathbb{E}_{x,z}[(D(G(x,z)) - a)^2] \quad (6)$$
$$L_{LScGAN(G)} = \frac{1}{2}\sum \mathbb{E}_{x,z}[(D(G(x,z)) - c)^2] \quad (7)$$

Where $a$ and $b$ are the labels for fake and real data, and $c$ is the value that $G$ wants $D$ to believe for fake data.

Although LSGANs address the gradient vanishing as well as training instability issues, they only consider pixel-wise differences between feature map $D(x)/\ D(G(x,z)$ and its corresponding label. On the contrary, SSIM considers the changes in overall structural information between the feature map and its target, providing a more holistic comparison and improved perceptual quality of reconstructed images. Taking another step forward, MS-SSIM generalizes single-scale SSIM to incorporate the variations of image resolution and viewing conditions[23]. Therefore, we propose to extend the objective of LSGANs with the addition of MS-SSIM for training of Re-Con-GAN. The loss function is designed as **Equation (8-9)**.

$$L_{LScGAN(D)} = \frac{1}{2}\left[\sum \mathbb{E}_{x,y}[(D(x,y) - b)^2] + \left(1 - MSSSIM(D(x,y), b)\right)\right] + \left[\frac{1}{2}\sum \mathbb{E}_{x,z}[(D(G(x,z) - a)^2] + \left(1 - MSSSIM(D(G(x,z), a)\right)\right] \quad (8)$$
$$L_{LScGAN(G)} = \frac{1}{2}\sum \mathbb{E}_{x,z}[(D(G(x,z) - c)^2]\left(1 - MSSSIM(D(G(x,z), c)\right) \quad (9)$$

Additionally, previous studies have demonstrated that mixing the GAN objective with a more traditional loss, such as $L_2$ or $L_1$ distances is more beneficial to the convergence of the generator. Both $L_2$ and $L_1$ distances have been explored by pioneers with $L_1$ distance proved to encourage less blurring over $L_2$[29]. Thus, our final objective for $L_{LScGAN(G)}$ is designed as **Equation (10-11)**.

$$L_{LScGAN(G)} = \frac{1}{2}\sum \mathbb{E}_{x,z}[(D(G(x,z) - c)^2]\left(1 - MSSSIM(D(G(x,z), c)\right) + \lambda L_{L_1}(G) \quad (10)$$
$$L_{L_1}(G) = \mathbb{E}_{x,y,z}[||G(x,z) - y||_1] \quad (11)$$

Where $\lambda$ is a hyperparameter and is set as 1 across all training.

### 2.2.3 Model Training

The pipeline was implemented with Pytorch, and the training was performed on a GPU workstation with $2 \times RTX\ 4090$. All the models were trained for 200 epochs, with the gradient linearly decayed after epoch 100. Adam optimizer with a learning rate of 0.0002 and batch size of $2 \times 4$ was applied.

## 2.3 Baseline Algorithms

Conventional CS and non-GAN DL approaches were included as benchmarks. DL baselines consist of 3D UNet (U256)[38], 9 blocks ResNet (ResNet9) [37], and RST-T[21].

Ablation studies that solely tune generators (ResNet9, U256 and RST-T) without generative adversarial training are also conducted to underpin the improvement made by Re-Con-GAN.

## 2.4 Evaluation

Image evaluation consists of two parts. First, we performed quantitative quality assessment against fully sample nnFFT using the following metrics: root mean squared error (RMSE), PSNR, SSIM, and inference time, shown in **Equation (12-14)**.

$$RMSE = \sqrt{\frac{\sum_{i=1}^{N}(G(x,z)-y)^2}{N}} \tag{12}$$

$$PSNR = 20 \cdot \log_{10}\frac{MAX_I}{RMSE} \tag{13}$$

$$SSIM = \frac{(2\mu_{G(x,z)}\mu_y+c_1)(2\sigma_{G(x,z)y}+c_2)}{(\mu_{G(x,z)}^2+\mu_y^2+c_1)(\sigma_{G(x,z)}^2+\sigma_y^2+c_2)} \tag{14}$$

Where $MAX_I$ is the max possible pixel value in a tensor, $\mu_{G(x,z)}$ and $\mu_y$ is the pixel mean of $G(x,z)$ and $y$ and $\sigma_{G(x,z)y}$ is the covariance between $G(x,z)$ and $y$, $\sigma_{G(x,z)}^2$ and $\sigma_y^2$ is the variance of $G(x,z)$ and $y$. Lastly, $c_1 = (k_1L)^2$ and $c_2 = (k_2L)^2$, where $k_1 = 0.01$ and $k_2 = 0.03$ in the current work and $L$ is the dynamic range of the pixel values ($2^{\#\ bits\ per\ pixel} - 1$).

Second, a radiotherapy-specific task was performed to test the accuracy of liver tumor detection and segmentation using an automated liver tumor segmentation network trained on a separate static 3D MR data cohort with a similar imaging protocol. Specifically, 70 patients (excluding the 48 4D MR data cohort) containing 103 manual GTV contours were scanned on the same 3T MRI scanner (MAGNETOM Vida, Siemens Healthcare, Erlangen, Germany) after injection of hepatobiliary contrast (gadoxetic acid; Eovist, Bayer) for each patient. A prototype free-breathing T1-weighted volumetric Cartesian sequence was used for 3D MRI acquisition. The scanning parameters were – TE=1.35 ms, TR=4.05 ms, matrix size = 260x320, in-plane resolution=1.09 mm × 1.09 mm and slice thickness=3 mm.

Mask-RCNN[34] has been used for various types of tumor detection and segmentation[42–44]. We employed this framework for the current task of liver GTV detection+segmentation with reconstructed accelerated images. ResNet50[45] without weight-frozen per training stage was used as backbone. ImageNet[46] pretrained weights followed by 3D static MR dataset fine-tuning was implemented for yielding the model convergence. Both mask as well as detection heads were turned on during network training[34]. The pipeline was implemented with Pytorch, and the training was performed on a GPU workstation with $4 \times RTX\ A6000$. All the models were trained for 80k iterations, with a learning rate 0.02 which is decreated by 10 at the 50k and 70k iterations. Stochastic gradient descent optimizer with a batch size of 32 ($4 \times 8$), weight decay of 0.0001 and momentum of 0.9 was used. The final training loss decreased to ~0.03. The 70 3D MR patients were used as the training set for tuning the detection+segmentation network. Images with positive GTV annotations were 3 times augmented in the training set to balance the ratio between positive and negative images. The training data is geometrically augmented using random resizing (image largest width to 640-800), horizontal flipping (p=0.5), and random rotation (angle 0-180°) and morphologically augmented using random gaussian noise (p=0.5, kernel=5, sigma=1) and random brightness (p=0.5).

The trained network segmented liver tumor in the 3x, 6x, and 10x images from the validation set of 4D MR (11 patients; 14 GTVs) processed by Re-Con-GAN, U256, ResNet9, RST-T, CS along with FS nuFFT and 3x, 6x and 10x US nuFFT validation images. Since the detection+segmentation network was designed to detect region of interest in 2D, we ignored the inter-z-dimension and inter-temporal-dimension relationship and organized all the images from 3D training and 4D test sets as 2D frames. All the images were z-score normalized, black border cropped out, and resized to $512 \times 512$. Both images with and without positive GTV annotations were included during the training and test stages.

Image-wise object detection (intersection over union threshold=0.5) and segmentation precision, recall and Dice score as well as 2D segmentation 95% Hausdorff distance ( $d_{H95}$) were used to evaluate the model performance as shown in **Equation (15-18)**.

$$Precision = \frac{TP}{TP+FP} \quad (15)$$

$$Recall = \frac{TP}{TP+FN} \quad (16)$$

$$Dice = \frac{TP}{TP+\frac{1}{2}(FP+FN)} \quad (17)$$

$$d_{H95}(X,Y) = MAX_{95}[d_{XY}, d_{YX}] = MAX_{95}\left[MAX_{95,x\in X} MIN_{95,y\in Y} d(x,y), MAX_{95,y\in Y} MIN_{95,x\in X} d(x,y)\right] \quad (18)$$

Where TP, FP and FN stand for true positive, false positive and false negative, $MAX_{95}$ and $MIN_{95}$ represents the 95th percentile of the distances between boundary points in $X$ and $Y$.

## 3 Experiments

Reconstruction statistical results and visualization of the validation set are reported in Table 1, **Figure 4,** and **Figure 5**.

Visually, **Figure 4** shows that as the acceleration ratio increases from 3x to 10x and the under-sampled nuFFT input degraded, Re-Con-GAN architectures gradually lost prediction sharpness while showing increasing streaking and tiling artifacts. Nonetheless, Re-Con-GAN with ResNet9 and U256 generators recovered sharper and more detailed morphologies than RST-T. Regarding the quantitative metrics of Re-Con-GAN, the architecture with the ResNet9 generator performed slightly better than that with the U256 generator, while the prediction of Re-Con-GAN with RST-T generator vastly degraded in comparison to the other two. Two different loss objectives ($L_1 + L_2$ and $L_1 + L_2$ +MS-SSIM) were compared during Re-Con-GAN training, with the addition of MS-SSIM encouraging slightly better model convergence. The per-patient inference speed of Re-Con-GAN with ResNet9 and U256 is 150 ms and 160 ms, respectively, meeting the requirements of real-time 4D MR reconstruction (<500 ms)[47].

Re-Con-GAN with the ResNet9 generator slightly outperformed CS, which is comparable to Re-Con-GAN with the U256 generator and substantially better than Re-Con-GAN with RST-T generator. As shown in Figure 4 and Figure 5**,** CS reconstruction results for 3x and 6x acceleration show minimal artifacts and good detail retention. CS shows more obvious streaking artifacts than Re-Con-Gan when increasing the acceleration to 10x. CS reconstruction time of 120 s is ~700X longer than Re-Con-Gan.

GTV detection and segmentation statistical results and visualization of an example from the validation set are reported in Table 2 and Figure 6. Generally speaking, the liver tumor can be reliably segmented using images acquired with up to 5x acceleration, but the performance dropped sharply with 10x. All the images reconstructed from different models (proposed and benchmarks) can, to varying degrees, improve the detection and segmentation results than US nuFFT images. From Table 2, we can observe that Re-Con-GAN ResNet9 achieved slightly inferior outcomes than FS nuFFT but was consistently superior to other benchmarks, including CS, Re-Con-GAN with U256 and RST-T generators and 3D non-adversarial trained networks. All models achieved better precision than recall, indicating a systematic under-segmentation/detection using the network.

From Figure 6, we can see that the GTV was still detectable at a 100% confidence score on a 3x US nuFFT image frame, where the confidence score dropped to 79% on the 6x frame, and the model completely missed its target on the 10x frame. Mask-RCNN can accurately detect and segment GTV across all acceleration levels on Re-Con-GAN and CS reconstructed images, while Re-Con-GAN achieved a moderately higher confidence score (98%) than that of CS (90%) at 10x acceleration.

| Model | Generator | Acceleration | Loss | PSNR↑ | 1-SSIM↓ | RMSE↓ | Time (s)↓ |
|---|---|---|---|---|---|---|---|
| Re-Con-GAN | ResNet9 | 3x | $L_1 + L_2$ | $25.65 \pm 2.89$ | $0.06 \pm 0.02$ | $0.09 \pm 0.03$ | **0.15** |
| | | | $L_1 + L_2$ +MS-SSIM | $\mathbf{26.13 \pm 3.02}$ | $\mathbf{0.05 \pm 0.02}$ | $\mathbf{0.08 \pm 0.03}$ | |
| | | 6x | $L_1 + L_2$ | $21.68 \pm 2.88$ | $0.10 \pm 0.03$ | $0.13 \pm 0.03$ | |
| | | | $L_1 + L_2$ +MS-SSIM | $\mathbf{23.97 \pm 3.84}$ | $\mathbf{0.07 \pm 0.03}$ | $\mathbf{0.11 \pm 0.04}$ | |
| | | 10x | $L_1 + L_2$ | $20.01 \pm 2.81$ | $0.11 \pm 0.03$ | $0.15 \pm 0.05$ | |
| | | | $L_1 + L_2$ +MS-SSIM | $\mathbf{21.61 \pm 2.93}$ | $\mathbf{0.09 \pm 0.03}$ | $\mathbf{0.13 \pm 0.04}$ | |
| | U256 | 3x | $L_1 + L_2$ | $25.09 \pm 2.74$ | $0.10 \pm 0.03$ | $0.10 \pm 0.03$ | 0.16 |
| | | | $L_1 + L_2$ +MS-SSIM | $25.41 \pm 2.70$ | $0.08 \pm 0.03$ | $0.09 \pm 0.02$ | |
| | | 6x | $L_1 + L_2$ | $21.82 \pm 3.12$ | $0.10 \pm 0.04$ | $0.13 \pm 0.04$ | |
| | | | $L_1 + L_2$ +MS-SSIM | $22.01 \pm 3.13$ | $0.08 \pm 0.03$ | $0.12 \pm 0.04$ | |
| | | 10x | $L_1 + L_2$ | $19.95 \pm 3.01$ | $0.12 \pm 0.03$ | $0.14 \pm 0.05$ | |
| | | | $L_1 + L_2$ +MS-SSIM | $20.08 \pm 2.99$ | $0.11 \pm 0.03$ | $0.12 \pm 0.05$ | |
| | SWT-T | 3x | $L_1 + L_2$ | $19.22 \pm 2.64$ | $0.18 \pm 0.07$ | $0.18 \pm 0.11$ | 0.73 |
| | | | $L_1 + L_2$ +MS-SSIM | $20.21 \pm 2.76$ | $0.16 \pm 0.07$ | $0.15 \pm 0.09$ | |
| | | 6x | $L_1 + L_2$ | $18.05 \pm 3.11$ | $0.21 \pm 0.11$ | $0.20 \pm 0.13$ | |
| | | | $L_1 + L_2$ +MS-SSIM | $18.85 \pm 3.14$ | $0.21 \pm 0.11$ | $0.19 \pm 0.12$ | |
| | | 10x | $L_1 + L_2$ | $15.78 \pm 3.09$ | $0.28 \pm 0.10$ | $0.24 \pm 0.14$ | |
| | | | $L_1 + L_2$ +MS-SSIM | $15.97 \pm 3.01$ | $0.27 \pm 0.09$ | $0.23 \pm 0.13$ | |
| U256 | - | 3x | $L_1 + L_2$ +MS-SSIM | $22.23 \pm 2.95$ | $0.12 \pm 0.04$ | $0.13 \pm 0.05$ | 0.16 |
| | | 6x | | $19.02 \pm 3.12$ | $0.13 \pm 0.05$ | $0.16 \pm 0.06$ | |
| | | 10x | | $17.34 \pm 2.95$ | $0.15 \pm 0.05$ | $0.19 \pm 0.06$ | |
| SWT-T | | 3x | | $18.91 \pm 2.81$ | $0.21 \pm 0.09$ | $0.20 \pm 0.09$ | 0.73 |
| | | 6x | | $18.08 \pm 2.95$ | $0.24 \pm 0.08$ | $0.22 \pm 0.12$ | |
| | | 10x | | $14.52 \pm 3.11$ | $0.29 \pm 0.12$ | $0.27 \pm 0.17$ | |
| ResNet9 | | 3x | | $22.45 \pm 3.01$ | $0.11 \pm 0.03$ | $0.12 \pm 0.04$ | 0.15 |
| | | 6x | | $20.08 \pm 3.12$ | $0.12 \pm 0.04$ | $0.14 \pm 0.05$ | |
| | | 10x | | $18.25 \pm 3.10$ | $0.14 \pm 0.06$ | $0.17 \pm 0.06$ | |
| CS | | 3x | - | $25.31 \pm 2.56$ | $0.08 \pm 0.02$ | $0.19 \pm 0.05$ | 120 |
| | | 6x | | $20.73 \pm 2.95$ | $0.12 \pm 0.03$ | $0.16 \pm 0.05$ | |
| | | 10x | | $19.29 \pm 2.99$ | $0.13 \pm 0.05$ | $0.21 \pm 0.08$ | |

**Table 1**: Statistical results from our proposed Re-Con-GAN under 3x, 6x and 10x acceleration rate and their corresponding baselines are presented. The best score and the worst score under each acceleration is bolded and wavy underlined, respectively. The up arrows next the evaluation metrics means that a higher value is superior and vice-versa for the down arrow. All the statistics are calculated with images normalized to [0,1] scale.

| | | | Detection | | | Segmentation | | | |
|---|---|---|---|---|---|---|---|---|---|
| Image Modality | Generator | Acceleration | Precision↑ | Recall↑ | $Dice$ ↑ | Precision↑ | Recall↑ | $Dice$ ↑ | HD 95↓ (mm) |
| FS nuFFT | - | - | **94.40** | **72.91** | **82.27** | **92.52** | **72.55** | **81.33** | **8.87** |
| US nuFFT | - | 3x | 80.27 | 61.45 | 69.61 | 76.35 | 57.23 | 65.42 | 13.29 |
| | | 6x | 55.34 | 45.27 | 49.80 | 49.37 | 40.56 | 44.53 | 18.79 |
| | | 10x | 25.34 | 17.21 | 20.50 | 21.56 | 13.75 | 16.79 | 19.31 |
| Re-Con-GAN | ResNet9 | 3x | **93.57** | **71.38** | **80.98** | **91.26** | **70.07** | **79.27** | **8.95** |
| | | 6x | **91.05** | **70.32** | **79.35** | **89.77** | **68.38** | **77.63** | **9.13** |
| | | 10x | **85.45** | **67.46** | **75.40** | **82.06** | **62.57** | **71.00** | **9.27** |
| | U256 | 3x | 92.47 | 70.35 | 79.91 | 89.46 | 68.32 | 77.47 | 9.07 |
| | | 6x | 87.78 | 68.25 | 76.79 | 86.87 | 65.34 | 74.58 | 9.18 |
| | | 10x | 81.46 | 64.57 | 72.04 | 79.34 | 60.54 | 68.68 | 9.47 |
| | SWT-T | 3x | 82.74 | 61.36 | 70.46 | 78.35 | 59.37 | 67.55 | 12.89 |
| | | 6x | 78.35 | 57.23 | 66.15 | 75.47 | 53.27 | 62.46 | 14.73 |
| | | 10x | 60.45 | 49.46 | 54.41 | 59.48 | 48.57 | 53.47 | 16.81 |
| U256 | - | 3x | 92.35 | 70.37 | 79.88 | 89.02 | 67.99 | 77.10 | 9.10 |
| | | 6x | 86.89 | 68.12 | 76.37 | 86.08 | 65.24 | 74.22 | 9.25 |
| | | 10x | 81.01 | 63.75 | 71.35 | 78.99 | 60.12 | 68.28 | 9.90 |
| SWT-T | - | 3x | 81.35 | 60.12 | 69.14 | 77.24 | 59.01 | 66.91 | 13.52 |
| | | 6x | 77.45 | 56.34 | 65.23 | 75.06 | 53.17 | 62.25 | 15.04 |
| | | 10x | 68.72 | 54.36 | 60.70 | 58.77 | 47.62 | 52.61 | 17.08 |
| ResNet9 | - | 3x | 93.06 | 71.12 | 80.62 | 91.05 | 69.57 | 78.87 | 9.02 |
| | | 6x | 90.05 | 69.23 | 78.28 | 88.01 | 67.03 | 76.10 | 9.12 |
| | | 10x | 83.53 | 66.89 | 74.29 | 80.27 | 61.05 | 69.35 | 9.37 |
| CS | - | 3x | 93.46 | 71.06 | 80.74 | 91.11 | 70.02 | 79.18 | 8.99 |
| | | 6x | 90.89 | 70.01 | 79.10 | 88.72 | 67.33 | 76.56 | 9.12 |
| | | 10x | 84.56 | 67.02 | 74.78 | 81.99 | 62.04 | 70.63 | 9.35 |

**Table 2**: Statistical results from Mask-RCNN detection and segmentation from our proposed Re-Con-GAN under 3x, 6x and 10x acceleration rate and their corresponding baselines are presented. The best score and the worst score under each acceleration are bolded and wavy underlined, respectively. The up arrows next to the evaluation metrics mean that a higher value is superior and vice-versa for the down arrow.

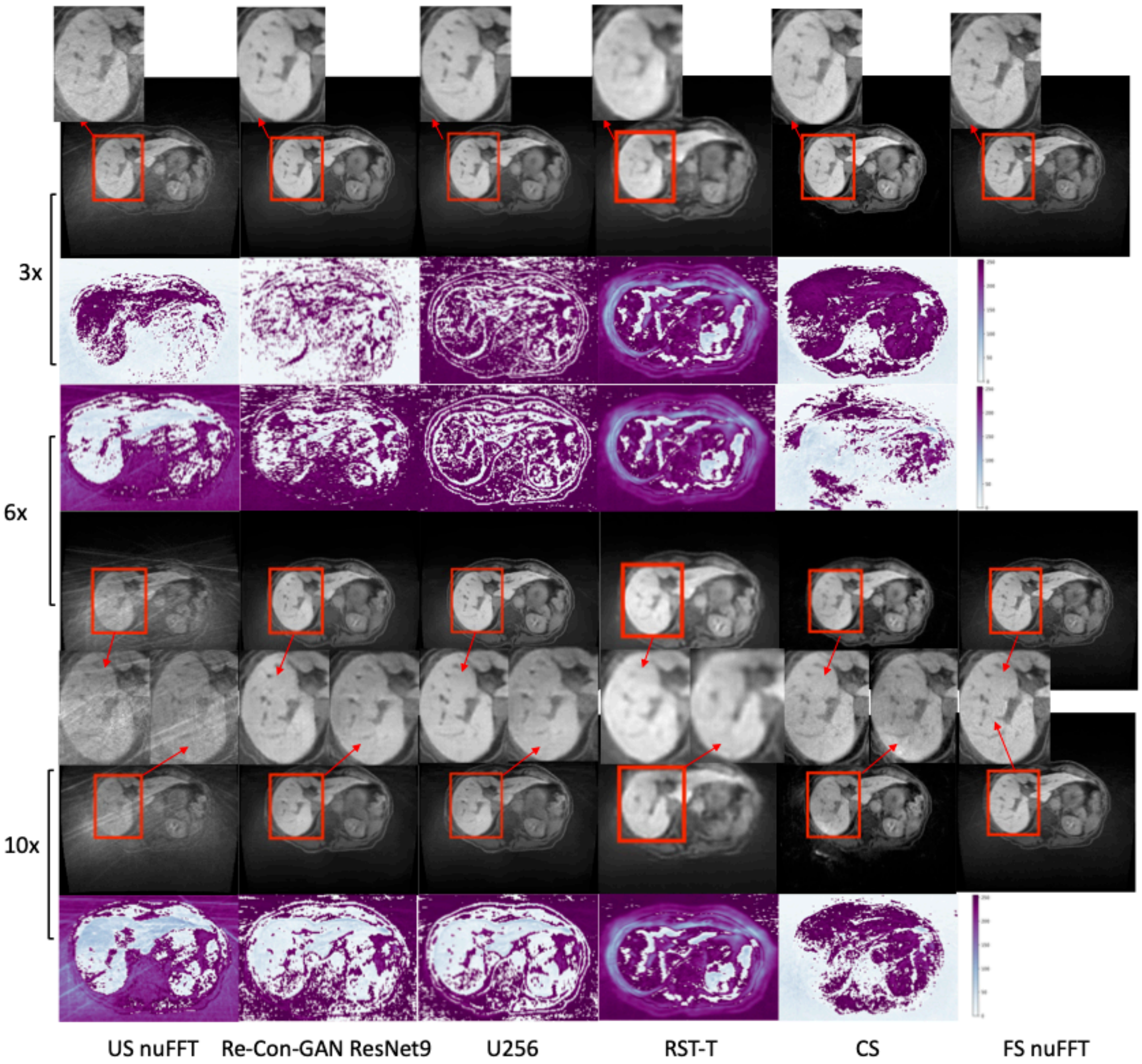

**Figure 4**: Visualization of 3x, 6x and 10x reconstruction results of an axial view slice from a patient in validation set. Reconstruction visualization, zoomed-in region of interest as well as residual between prediction and the fully sampled nuFFT reconstructed image are visualized. The regions in the red boxes are magnified for visualization. Residual maps are black-border-cropped for visualization clarity. US nuFFT refers to the under-sampled nuFFT reconstructed image series (input), Re-Con-GAN ResNet9 refers to the Re-Con-GAN with ResNet9 generator reconstruction result, U256 refers to the 3D UNet reconstruction result, RST-T refers to the 3D RST-T reconstruction result, CS refers to the compressed sensing reconstruction result, and FS nuFFT refers to the fully sampled nuFFT reconstructed images (GT). All the images are visualized after normalizing to [0,1] scale.

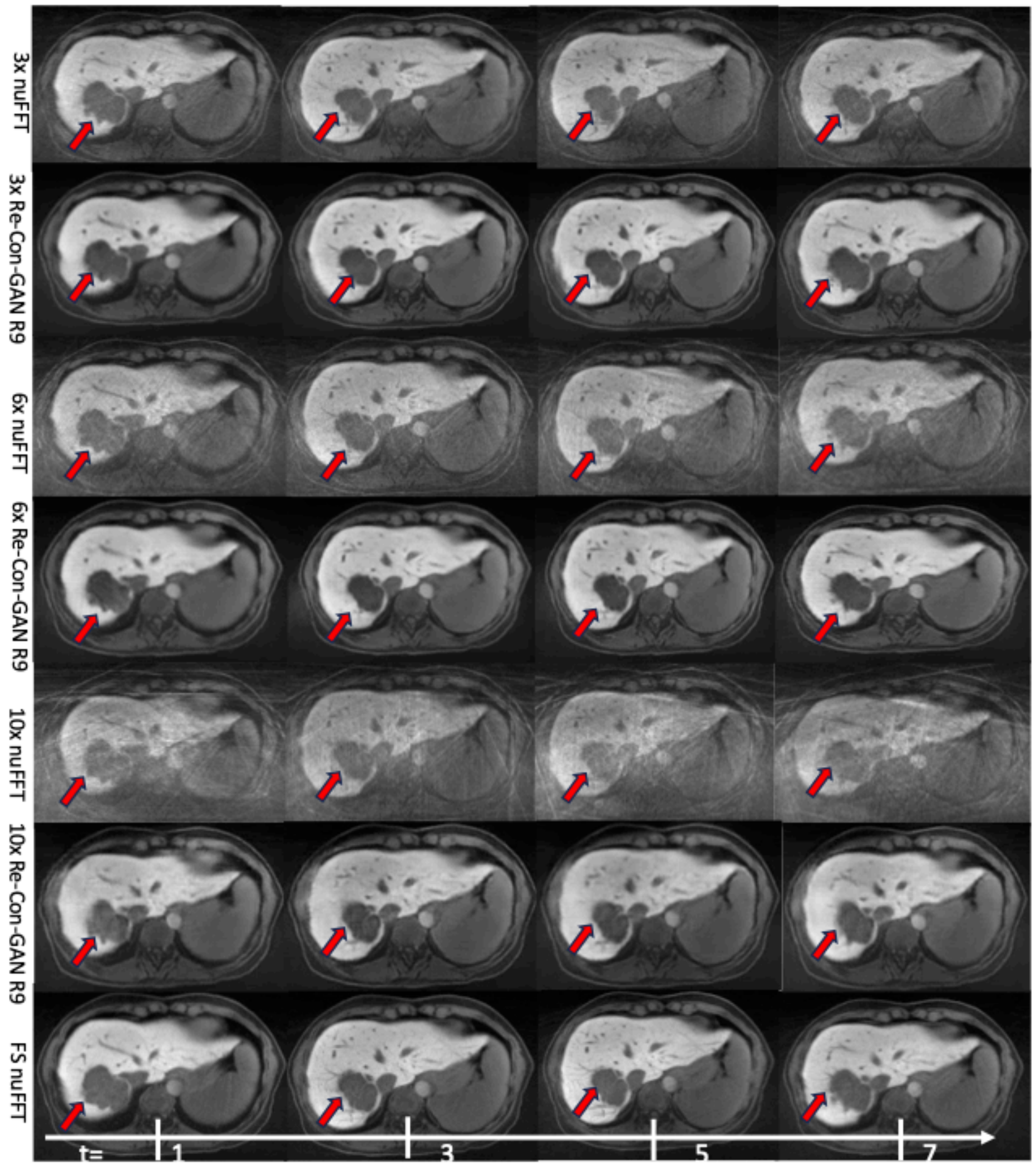

**Figure 5**: Visualization of a selected temporal profile (motion binning = 1, 3, 5, 7) from a patient in the validation set. 3x, 6x, and 10x reconstruction results from input, GT and our proposed method are visualized. Red arrows denote the patient's GTV. All the images are visualized after normalizing to [0,1] scale.

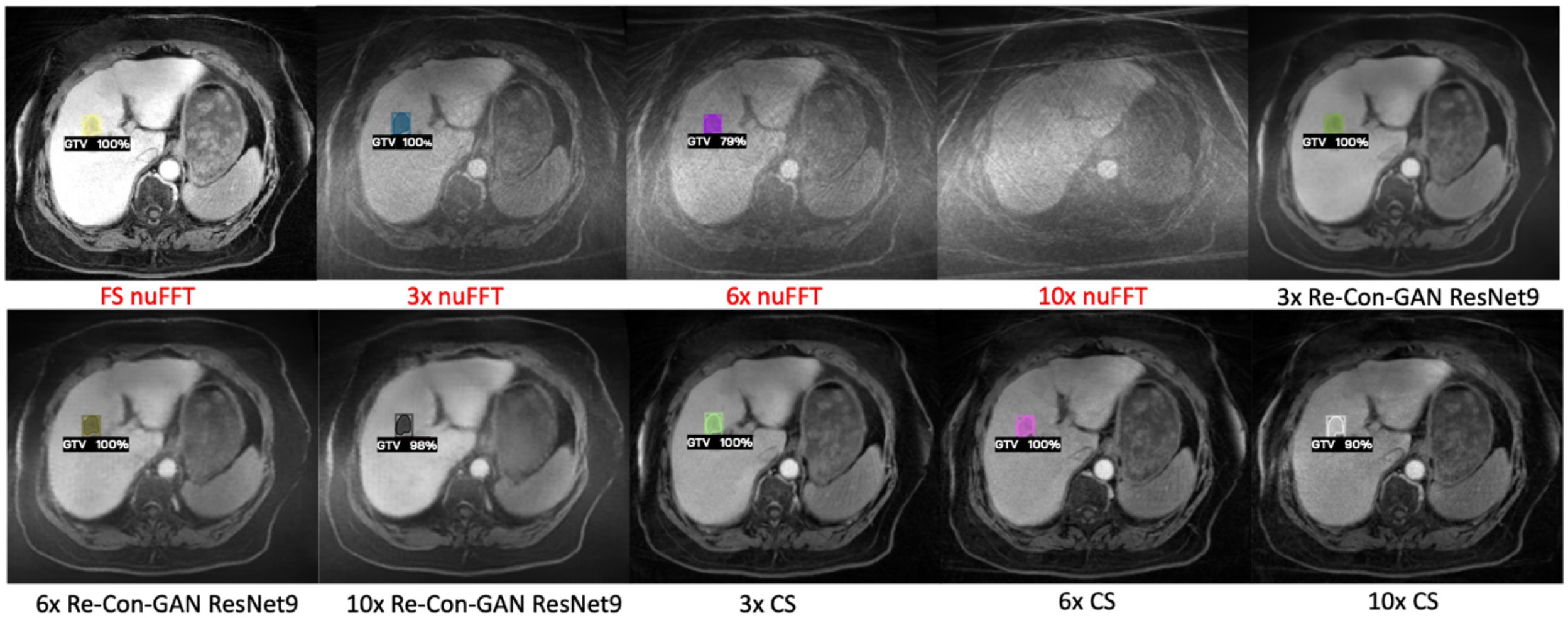

**Figure 6**: Visualization of Mask-RCNN detection and segmentation results on a validation set patient from selected Re-Con-GAN and baseline models. The detection bounding box and segmentation mask made by Mask-RCNN are visualized on top of its corresponding input image modality with confidence scores. All the images are visualized after normalizing to [0,1] scale.

## 4 Discussion

The paper focuses on liver 4D MRI, which is particularly relevant to image-guided liver cancer radiotherapy. Hepatocellular carcinoma (HCC) is the fifth most common cancer worldwide in men and the seventh in women. HCC represents the third most frequent and fast-rising cause of cancer deaths[48,49]. The past few decades have witnessed a continuous decrease in the average age at HCC diagnosis, with most HCC patients now diagnosed between 45 and 60[50]. Additionally, the liver is one of the most common metastatic sites of several cancer types, including colorectal, pancreatic, melanoma, lung, and breast cancer[51].

Surgical resection remains the standard of care for hepatic primary and metastatic tumors and continues to demonstrate persistent positive prognosis outcomes in surgical-qualified candidates[52]. For non-surgical patients, orthotopic liver transplants, ablative procedures, chemotherapy, and radiation therapy (RT) are considered effective alternatives[53]. Stereotactic body radiation therapy (SBRT), delivering intense and highly conformal radiation doses, has shown promising results in hepatic malignancy and metastasis management[53–56]. The success of liver SBRT, however, depends on the ability to focus the high radiation dose on the tumor while minimizing the dose to the normal liver tissue, which is sensitive to radiation[57]. A prerequisite for successful liver SBRT is accurate liver tumor imaging and motion management of the highly mobile organ.

Unlike lung tumors, which are often clearly visualized in Computed Tomography (CT) and 4D CT, liver tumors have low soft tissue X-ray contrast but high MR contrast, making MRI and 4D MRI an ideal pre and during-treatment liver imaging. 4D MRI requires densely sampled k-space data for spatiotemporal reconstruction. Fully sampling the required k-space data results in lengthy MR sequences that are challenging for MR simulation due to limited patient tolerance and available scanner time and impractical for online MR-guided RT[58]. Acceleration of MR acquisition via down-sampling the k-space and rapid image reconstruction without compromising the usability of the image quality is thus highly desired. Non-cartesian k-space sampling and compressed sensing have achieved remarkable success in the former goal but struggled with the latter due to the slow iterative algorithms. Though some previous works attempted to utilize DL methods, such as 3D UNet, RNN, and Transformers variants[16,20,21], to tackle the problem, these methods are heterogeneous in reconstructing dynamic liver images, as shown in the non-adversarial trained DL benchmarking results in **Figure 4** and **Table 1**. We postulate that the difficulty of defining a loss function suitable for simultaneous detail retention and artifacts suppression is a contributing factor.

Therefore, we developed Re-Con-GAN in this work. Re-Con-GAN is structured with pair-trained conditional GAN architecture constraint with loss objective fused from $L_2$, $L_1$ and MS-SSIM. Three types of generators, including 3D ResNet9, 3D UNet and 3D RST-T, are demonstrated. $70 \times 70$ PatchGAN is utilized as the discriminator. Re-Con-GAN is validated on an in-house dynamic liver MRI dataset with 48 patients having a total of 12332 2D+t image series. Further downstream validation tasks of GTV detection and segmentation were also conducted on Re-Con-GAN reconstructed images. Re-Con-GAN showed competitive performance to CS in image quality and is significantly faster. The real-time inference speed and sharp liver GTV morphology visualization of Re-Con-GAN are conducive to image-guided liver radiotherapy. Our proposed methods achieve 1-SSIM of $0.05 \pm 0.02$ at 3x acceleration, which outperforms previous GAN-based 3D stack-of-radial liver MRI reconstruction studies conducted by Gao et al. reporting 1-SSIM of $0.16 \pm 0.01$ at 3x acceleration[33]. The study based on raw k-space data and undersampling radial spokes of the stack of stars can be readily deployed.

There are several theoretical and practical advantages to using GAN for 4D MRI. Standard NNs, such as U256, ResNet9, and RST-T, fully parameterize their loss function and use the fixed loss function to conduct representation learning from training information. In GANs, the penalty imposed by the discriminator is a nonparametric loss function, mitigating the inflexibility of an explicitly defined loss function and tradeoffs in noise, uniformity, detail retention, and computational tractability. As shown here, Re-Con-GAN reconstructs sharper and more consistent images than the compared DL benchmarks (U256, ResNet9, and RST-T). The improvement is more evident in quantitative image quality assessment using SSIM, PSNR, and RMSE than in the automated liver segmentation task. Liver segmentation using deep learning is less sensitive to image quality but more dependent on the training data size, which is the common bottleneck of the current study. This is evidenced by a larger improvement in the segmentation accuracy with a higher acceleration ratio, where the image quality degradation is evident. We also note that the All generators significantly outperformed Transformers (RST-T). Among all the compared generators (U256, ResNet9, and RST-T), CNN architectures achieve similar performance, with U256 slightly inferior to ResNet9. We attribute the result to the current limited size of training samples. Evidence has shown that Vision Transformers architecture performance declines when trained on small datasets due to the lack of locality, inductive biases, and hierarchical structure of the representation commonly observed in CNNs. Therefore, Vision Transformers architectures, including RST-T, require large-scale training data or domain-relevant pre-training + fine-tuning to learn such properties from the data distribution[59].

The current work can be improved or extended in several areas. First, our implementation is restricted to learning 2D+t image series. 3D+t training would allow more effective learning of the inter-slice anatomy but requires an exceedingly large GPU memory footprint. Second, the current validation is conducted on a dataset collected from a single institute. Although our pipeline is shown robust to the single institutional held-out test, its performance in the external data needs further testing. Despite the recent rapid increase of medical images in the public domain, raw k-space data of 4D MRI essential for realistic undersampling are rarely stored and shared. Third, as more aggressive acceleration ratios (6x and 10x) are pursued, tiling artifacts was suppressed but still noticeable. Model structures more robust to tiling artifacts, such as diffusion-based frameworks[60] or post-processing techniques, are worth exploring to combat such artifacts. Fourth, despite the real-time image reconstruction speed, acquiring the highly under-sampled stack of star k-space data is not real-time. As a result, 4D MRI using Re-Con-GAN does not reflect real-time anatomy. Sparser sampling in combination with prior retrospective 4D MRI may be necessary for real-time 3D MR reconstruction. Lastly, the current method requires transformation from k-space to images as input. The additional step leads to information loss and added latency. Future work will explore networks using k-space or coil data as the input.

## 5 Conclusion

An efficient yet robust liver 4D MRI reconstruction framework, Re-Con-GAN, is proposed. Re-Con-GAN uses a flexible framework with 3D ResNet9, 3D UNet and 3D RST demonstrated as generator, PatchGAN as discriminator, and $L_1$, $L_2$ and MS-SSIM fused measurements as loss objectives. Validation from the in-house liver 4D MRI dataset substantiates the superior inference speed of Re-Con-GAN to its CS benchmark as well as higher predicted image quality to the compared 3D UNet, 3D ResNet and 3D RST-T DL Benchmarks.